\documentclass[12pt]{article}
\newcommand{\nin}{\noindent}
\newcommand{\be}{\begin{equation}}
\newcommand{\ee}{\end{equation}}
\newcommand{\bea}{\begin{eqnarray}}
\newcommand{\eea}{\end{eqnarray}}

\newcommand{\hf}{\frac{1}{2}}
\newcommand{\nn}{\nonumber\\}

\usepackage{graphicx}
                                                                                                                        
\begin{document}
                                                                                                                        
\begin{center}

{\Large{\bf A note on Liouville theory}}

\vspace{.2cm}

{\bf Jean Alexandre}

\vspace{.2cm}

Department of Physics, King's College, 

The Strand, London WC2R 2LS, U.K.

jean.alexandre@kcl.ac.uk
                                                                                                                        
\vspace{3cm}

{\bf Abstract}

\end{center}

An exact differential equation is derived for the evolution of the Liouville effective action 
with the mass parameter. This derivation is based on properties of the exponential potential and some 
consequences of the equation are discussed.

\vspace{3cm}

Liouville theory has been studied for a long time \cite{liouville} and is currently relevant to
two-dimensional quantum gravity or non critical string theory. 
For a recent review on Liouville formalism, see \cite{nakayama} and references therein. 
For a recent review on the phenomenological implications, see \cite{mavromatos} and references therein.

The classical action defining the theory is 
\be
S_\lambda[\tilde\phi]=\int d^2\xi \sqrt{g}\left\{\hf g_{ab}\partial^a\tilde\phi\partial^b\tilde\phi
+R\tilde\phi+\lambda m^2e^{\tilde\phi}\right\},
\ee
where $g_{ab}$ is the fixed world sheet metric, $g$ is its determinant and $R$ is the curvature scalar.
The dimensionless parameter $\lambda$ controls the strength of the only dimensionfull parameter $m^2$.
The field $\tilde\phi$ is dimensionless, leading to an infinite
set of classically marginal interactions. This is specific to the two-dimensional world on
which this field lives, which gives rise to the rich conformal properties of the model.

The classical Liouville theory is well known,
as well as the exact effective potential in a flat background \cite{jackiw}. Still, describing the full
quantum theory in a general background remains a difficult task.

The aim of this paper is to derive an exact evolution equation for the effective action (the one-particle
irreducible graphs generator functional) with $\lambda$ and to discuss some of its properties. 
We will see that this evolution equation (\ref{evolG}) has a linear structure and is a direct consequence 
of properties of the exponential potential.

\vspace{.5cm}

The partition function $Z_\lambda$ and the connected graphs generator functional $W_\lambda$ are defined by
\be
Z_\lambda[j]=e^{-W_\lambda[j]}=\int{\cal D}[\tilde\phi]\exp\left\{-S_\lambda[\tilde\phi]
-\int_\xi j\tilde\phi\right\},
\ee
where $j$ is the source and we use the notation
\be
\int_\xi(\cdot\cdot\cdot)=\int d^2\xi\sqrt{g(\xi)}(\cdot\cdot\cdot).
\ee 
In what follows, we consider a translation invariant measure for the path integral over the field $\tilde\phi$.
The quantum field $\phi$ is obtained by differentiation of $W_\lambda$: 
\be\label{1W}
\frac{1}{\sqrt{g(\xi)}}\frac{\delta W_\lambda[j]}{\delta j(\xi)}=
\frac{1}{Z_\lambda[j]}\left<\tilde\phi(\xi)\right>=\phi(\xi),
\ee
where 
\be
<(\cdot\cdot\cdot)>=\int{\cal D}[\tilde\phi](\cdot\cdot\cdot)
\exp\left\{-S_\lambda[\tilde\phi]-\int_\xi j\tilde\phi\right\} 
\ee
The one-particle irreducible graphs generator functional, the effective action $\Gamma_\lambda$, is
obtained by inverting the relation (\ref{1W}) between $\phi$ and $j$ and making the Legendre transform: 
\be
\Gamma_\lambda[\phi]=W_\lambda[j]-\int_\xi j\phi,
\ee
such that its functional derivative is
\be\label{1G}
\frac{1}{\sqrt{g(\xi)}}\frac{\delta\Gamma_\lambda[\phi]}{\delta\phi(\xi)}=-j(\xi).
\ee  

\vspace{.5cm}

The aim is to find the evolution of $\Gamma_\lambda$ with $\lambda$, and one
can obtain for this an exact differential equation for $\Gamma_\lambda$ which is now derived.
  
A change in the parameter $m^2$ amounts to a translation in the field $\tilde\phi$, such that
the partition function can be written
\be\label{symm}
Z_\lambda[j]=\int{\cal D}[\tilde\phi]\exp\left\{
-S_1[\ln\lambda+\tilde\phi]+\int_\xi R\ln\lambda-\int_\xi j\tilde\phi\right\}.
\ee
The change of variable $(\ln\lambda+\tilde\phi\to\tilde\phi)$ gives then
\bea
Z_\lambda[j]&=&\int{\cal D}[\tilde\phi]\exp\left\{
-S_1[\tilde\phi]-\int_\xi j\tilde\phi+\ln\lambda\int_\xi (R+j)\right\}\nn
&=&Z_1[j]\exp\left\{\ln\lambda\int_\xi(R+j)\right\}.
\eea
As a consequence, we have 
\be
\partial_\lambda Z_\lambda[j]=\frac{Z_1[j]}{\lambda}
\exp\left\{\ln\lambda\int_\xi(R+j)\right\}\int_\xi(R+j),
\ee
such that
\be\label{evolZ}
\lambda\partial_\lambda Z_\lambda[j]=Z_\lambda[j]\int_\xi(R+j).
\ee
To find the evolution of the effective action, one must keep in mind that its independent variables are
$\phi$ and $\lambda$ such that
\bea
\partial_\lambda\Gamma_\lambda[\phi]&=&\partial_\lambda W_\lambda[j]
+\int d^2\xi\frac{\delta W_\lambda[j]}{\delta j(\xi)}\partial_\lambda j(\xi)
-\int d^2\xi\sqrt{g(\xi)}\partial_\lambda j(\xi)\phi(\xi)\nn
&=&\partial_\lambda W_\lambda[j]=-\frac{\partial_\lambda Z_\lambda[j]}{Z_\lambda[j]}.
\eea
Using then Eqs.(\ref{evolZ}) and (\ref{1G}), we find 
\be\label{evolG}
\lambda\partial_\lambda\Gamma_\lambda[\phi]+4\pi\chi=\int d^2\xi~
\frac{\delta\Gamma_\lambda[\phi]}{\delta\phi(\xi)},
\ee
where 
\be
\chi=\frac{1}{4\pi}\int d^2\xi\sqrt{g}R 
\ee
is the Euler characteristic of the world sheet.

\vspace{.5cm}

An interesting point is that the evolution equation (\ref{evolG}) is linear in $\Gamma_\lambda$. As a 
consequence, starting from a given functional at $\lambda=0$, no new functional dependence can be 
generated by the flow in $\lambda$.
It can further be seen that a naive initial condition cannot be used to solve the 
evolution equation (\ref{evolG}): a "naive initial condition" would be $\Gamma_0=S_0$
since for $\lambda=0$ there are no interactions and thus the effective action must be the bare one.
But if this choice is made, the solution of
Eq.(\ref{evolG}) is $\Gamma_\lambda=S_\lambda$ for all values of $\lambda$ ($S_\lambda$ satisfies
Eq.(\ref{evolG}) up to a surface term). Thus the flow in $\lambda$ is discontinuous at $\lambda=0$.
This is another way to recover the well known fact that the Liouville theory is non perturbative,
and is consistent with the Wilson Renormalization Group approach given in \cite{wetterich},
where the authors explain that the initial (i.e. for large momentum) 
functional dependence of the running average action has to be chosen carefully, using Weyl-Ward
identities.

\vspace{.5cm}

One can derive constraints on the 1-particle-irreducible graphs of the theory, using the evolution 
equation (\ref{evolG}). These correlation functions are
\be
G^{(n)}_\lambda(\xi_1,\cdot\cdot\cdot,\xi_n)=\frac{\delta^n\Gamma_\lambda[\phi]}
{\delta\phi(\xi_1)\cdot\cdot\cdot\delta\phi(\xi_n)}|_{\phi=0},
\ee
and they satisfy, from Eq.(\ref{evolG}):
\be
\lambda\partial_\lambda G^{(n)}_\lambda(\xi_1,\cdot\cdot\cdot,\xi_n)=
\int d^2\xi~G^{(n+1)}_\lambda(\xi,\xi_1,\cdot\cdot\cdot,\xi_n)~~~~~~n\ge 1
\ee
These constraints are equivalent to some Ward identities based on the symmetry expressed by 
Eq.(\ref{symm}): a change in the mass parameter is equivalent to a translation in the field.
A set of sum rules relating $n$-point and $(n+1)$-point functions 
are also obtained in \cite{pawlowski} for correlation functions of 
vertex operators. In this work the authors do not deal with the Legendre transform $\Gamma$,
but use similar functional methods.

\vspace{.5cm}

The evolution equation (\ref{evolG}) actually leads to an interesting relation.
Define $\tilde\Gamma_t=\Gamma_\lambda+4\pi\chi t$, where $t=\ln\lambda$. Eq.(\ref{evolG})
reads then
\be
\partial_t\tilde\Gamma_t[\phi]=\int d^2\xi\frac{\delta\tilde\Gamma_t[\phi]}{\delta\phi(\xi)},
\ee
which leads to the following higher order derivatives:
\be
\frac{\partial^n\tilde\Gamma_t}{\partial t^n}[\phi]=
\int d^2\xi_1\cdot\cdot\cdot d^2\xi_n\frac{\delta^n\tilde\Gamma_t[\phi]}
{\delta\phi(\xi_1)\cdot\cdot\cdot\delta\phi(\xi_n)}.
\ee
Thus the ressumation can be written
\bea
\tilde\Gamma_t[\phi]&=&\sum_n \frac{t^n}{n!}\frac{\partial^n\tilde\Gamma_0}{\partial t^n}[\phi]\nn
&=&\sum_n \frac{t^n}{n!}\int d^2\xi_1\cdot\cdot\cdot d^2\xi_n\frac{\delta^n\tilde\Gamma_0[\phi]}
{\delta\phi(\xi_1)\cdot\cdot\cdot\delta\phi(\xi_n)}\nn
&=&\tilde\Gamma_0[t+\phi].
\eea
Finally, if we come back to the initial notations, we have
\be
\Gamma_\lambda[\phi]=\Gamma_1[\phi+\ln\lambda]-4\pi\chi\ln\lambda.
\ee 
More specifically, the effective potential $U_\lambda(\phi_0)$,
obtained for a constant configuration $\phi_0$, is of the form
\be
U_\lambda(\phi_0)=m^2u(\ln\lambda+\phi_0),
\ee
where $u$ is a dimensionless function. Thus the effective potential depends on the same combination 
$(\ln\lambda+\phi_0)$ as the bare potential.

\vspace{.5cm}
 
To conclude, Eq.(\ref{evolG}) has to be seen as complementary tool to understand the quantum structure
of Liouville theory. This evolution equation alone is not sufficient to generate the quantum theory:
to this end one needs in addition an initial condition at $\lambda=0$, which should
be set up using the conformal properties of the theory.

\vspace{.5cm}

\nin {\it Aknowledgement} I would like to thank N. Mavromatos for usefull suggestions.


\begin{thebibliography}{99}

\bibitem{liouville} J.Liouville, J.Math.Pures Appl.18 (1853) 71.

\bibitem{nakayama} Yu Nakayama, Int.J.Mod.Phys.A19 (2004) 2771.

\bibitem{mavromatos} J.Ellis, N.Mavromatos, D.V.Nanopoulos, gr-qc/0502119.

\bibitem{jackiw} E.D'Hoker, R.Jackiw, Phys.Rev.D26 (1982) 3517.

\bibitem{wetterich} M.Reuter, C.Wetterich, Nucl.Phys.B506 (1997) 483.

\bibitem{pawlowski} L.O'Raifeartaigh, J.M.Pawlowski, V.V.Sreedhar, Phys.Lett.B481 (2000) 436.

\end{thebibliography}
\end{document}